\documentclass[pra,twocolumn,notitlepage,floatfix,superscriptaddress]{revtex4-2}
\usepackage[T1]{fontenc}
\usepackage[colorlinks=true,urlcolor=blue,citecolor=blue,anchorcolor=red]{hyperref}
\usepackage{graphicx,url,color,physics,bbm,soul,mathtools,amssymb,amsmath,dsfont,lmodern}
\usepackage{mathtools}
\usepackage{braket}
\usepackage{notoccite}
\usepackage{dsfont}
\usepackage{circuitikz}
\usepackage{tikz}
\usepackage{soul}
\usepackage{graphicx}
\usetikzlibrary{quantikz}
\usepackage{tikz}
\usepackage[ruled]{algorithm2e}
\usepackage{multirow}
\usepackage{geometry}
\usepackage{amsthm}
\newtheorem{theorem}{Theorem}

\begin{document}
\newcommand{\orcidicon}[1]{\href{https://orcid.org/#1}{\includegraphics[height=\fontcharht\font`\B]{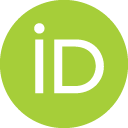}}}

\title{Phase Estimation with Compressed Controlled Time Evolution}

\author{Erenay Karacan\orcidicon{0009-0000-1399-5084}}
\email{ekaracan@ethz.ch}
\affiliation{Department of Physics, ETH Zurich, Otto-Stern-Weg 1,
8093 Zurich, Switzerland}
\date{\today}

\begin{abstract}
Many optimally scaling quantum simulation algorithms employ controlled time evolution of the Hamiltonian, which is typically the major bottleneck for their efficient implementation. This work establishes a compression protocol for encoding the controlled time evolution operator of translationally invariant, local Hamiltonians into a quantum circuit. It achieves a near-optimal  in time $t$  scaling  for  circuit depth  $\mathcal{O}(t  \text{ polylog}(t N/\epsilon))$, while reducing the control overhead from a multiplicative to an additive factor. We report that this compression protocol enables the implementation of Iterative Quantum Phase Estimation with as few as 414 CNOT gates for a frustrated quantum spin system on a 6$\times$6 triangular lattice and delivers ground state energy errors below 1\% (with $\pm$ 1.5\% variation, calculated with a hardware noise aware pipeline) on a 4$\times$4 triangular lattice using the noisy emulator of the Quantinuum H2 trapped ion device.
\end{abstract}

\maketitle

\section{Introduction}

Investigating ground state properties of strongly correlated quantum systems opens up new perspectives in understanding quantum matter. Exact, numerical simulations of such many-body systems are notoriously hard, as the Hilbert space dimension of these systems grows exponentially with system size.  Due to this difficulty of simulating quantum systems in classical computers, a new computational paradigm has emerged in the past decades following the idea of simulating quantum systems on quantum computers \cite{feynman82}.  Among many quantum protocols proposed for this purpose \cite{Dalzell_2025, babbush2025grandchallengequantumapplications, arnault2024typologyquantumalgorithms, Peruzzo_2014, McClean_2016, FaEtAl00, AlLi18, Low_2017, Low2019hamiltonian, farhi1996analoganaloguedigitalquantum, Lloyd1996}, one digital (gate-based) quantum protocol with well-defined error bounds, well-established scaling to large systems and high accuracy calculations, is the Quantum Phase Estimation (QPE) algorithm, where an input state is projected onto an eigen-manifold of the target Hamiltonian through multiple ancilla measurements \cite{kitaev1995quantummeasurementsabelianstabilizer, Cleve_1998}.

\begin{figure*}
    \centering
    \includegraphics[width=0.99\linewidth]{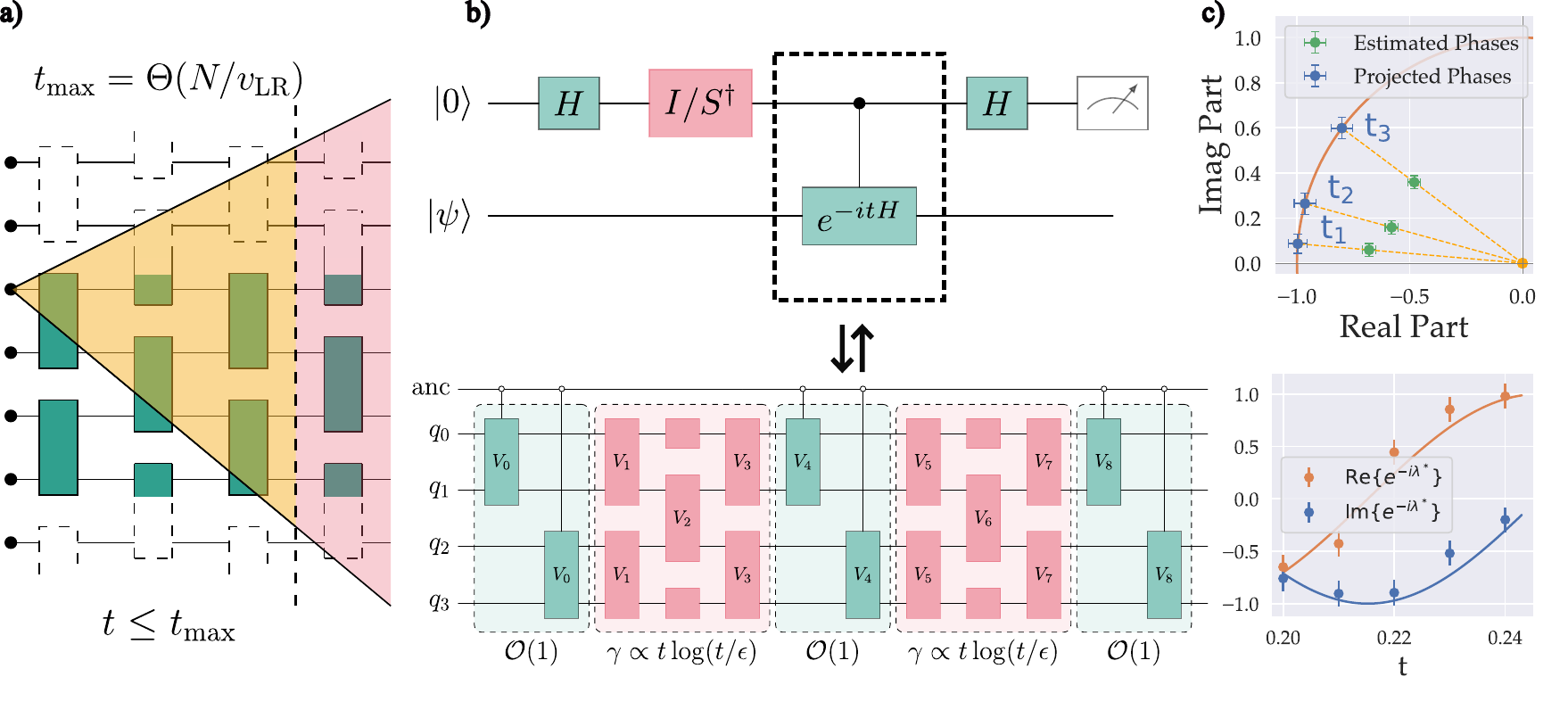}
    \caption{\textbf{Conceptual visualization of the proposed protocol. (a) } 
    Brickwall Ansatz of $N=4$ qubits, employed to approximate the time evolution dynamics of a local, translationally symmetric Hamiltonian in one dimension. The light-cone shaped trajectory illustrates the information propagation with a finite velocity $v_{\text{LR}}$. Gates, optimized on this Ansatz, can be reused to simulate time dynamics of larger systems, if the targeted evolution time does not exceed a $t_{\text{max}}=\mathcal{O}(N/v_{\text{LR}})$, because for $t<t_{\text{max}}$ the light cone doesn't expand enough to entangle sites $(X, Y)$ with distance $\hspace{0.1cm} \mathclap{l}\hspace{0.1cm}(X,Y) \ge N$. \textbf{(b)} \textbf{Top:} Iterative Quantum Phase Estimation circuit. By inserting identity (inverse phase gate) in the red block, one samples the real (imaginary) part of the phase $e^{-it \lambda}$. \textbf{Bottom:} Translationally Invariant Compressed Control (TICC) circuit that approximates the controlled time evolution for a chain of four qubits. In this circuit, green gates effectively flip the evolution direction, hence controlling these green gates with the ancilla implements the equivalence given in Eq. \ref{eq:cU_equivalence}. Crucially, the number of green layers does not scale with respect to the targeted evolution time. Number of red layers $\gamma$, scales near-optimally with respect to evolution time $t$. \textbf{(c)} Post processing after the Hadamard test. \textbf{Top:} renormalizing the estimated phase amplitude, amplifies statistical error bars by projection onto the unit circle, incorporating depolarizing noise effects. \textbf{Bottom:} fit of measured phases for several time points into a phase curve, from which we infer the phase angle estimation ($e^{-i t  E_{\text{fit}}}$).} 
    \label{fig:fig1}
\end{figure*}

The main limitation in implementing a large-scale QPE protocol on near-term quantum hardware is encoding the controlled time evolution operator of the target Hamiltonian into a quantum circuit. Not only for the QPE, but for many other optimally scaling quantum simulation algorithms (e.g. ground state preparation through eigenstate filtering \cite{qetu, aff, aqcf} and block encoding based methods \cite{Gily_n_2019, Low_2017, Kimmel_2017}), efficiently implementing the controlled time evolution operator constitutes a bottleneck. As such algorithms are highly sensitive to the approximation error of the controlled time evolution, gate counts needed for their high precision implementation on large systems easily exceed the coherence thresholds of current quantum devices. 

Although many proposals can efficiently compress the uncontrolled time evolution operator for various system classes of interest \cite{Tepaske_2023, Mansuroglu_2023, lin_compression, kokcu_2022, rqcopt1, rqcopt2}, none can avoid the overhead incurred when every two-qubit gate in the time evolution circuit must be promoted to a controlled operation. This overhead increases the circuit depth by a multiplicative factor, which we call \textit{control factor} and denote by $\gamma_D$. For an exact decomposition of an arbitrary controlled two-qubit gate into a sequence of uncontrolled two-qubit gates, $\gamma_D$ is typically in the range 15-18, as suggested by standard constructions \cite{Barenco_1995}. If one allows approximate decompositions, this factor can be reduced, at the cost of introducing a decomposition error $\epsilon_D$.

In this paper, we present a compression scheme for the controlled time evolution operator of translationally invariant (TI), local Hamiltonians, that both scales near-optimally with respect to evolution time and alleviates control overheads by reducing the control factor $\gamma_D$ from a multiplicative to an additive term. We benchmark the performance of our protocol by comparing it to other compression proposals and product formula based approaches, on various one- and two-dimensional quantum spin systems, as well as employ our compressed circuit in Quantum Phase Estimation protocols (both the iterative \cite{rpe} and standard variants \cite{kitaev1995quantummeasurementsabelianstabilizer, Cleve_1998}). Our QPE protocol is conceptually visualized in Fig. \ref{fig:fig1}. 

There is a well-established equivalence between a sequence where the ancilla controls the evolution with time $t$ and a sequence where the ancilla qubit acts as a control switch between $+t/2$ and $-t/2$ (forwards and backwards) evolutions \cite{Babbush_2018, qetu}. This equivalence is often exploited to reduce the multiplicative control overhead to an additive term. 

The key idea behind our approach is to leverage this equivalence \cite{Babbush_2018, qetu} in a circuit compression protocol, where the optimizer is made aware that the user will employ the compressed time evolution circuit later in a controlled evolution sequence.  We achieve this by splitting the cost function of the optimization into two additive parts, where one term is minimized when a subset of the circuit (e.g. red layers in Fig. \ref{fig:fig1}. b) approximates the backwards time evolution and the other term is minimized when the full circuit approximates the forwards time evolution. In this setting, one has to control only the complement of the aforementioned subset (e.g. green layers in Fig. \ref{fig:fig1}. b) with an ancilla to implement the globally controlled time evolution.  This way, we can reduce the control overhead to an additive term while keeping the favorable scaling behaviour of circuit compression methods.

\begin{table*}
    \centering
    \includegraphics[width=0.99\linewidth]{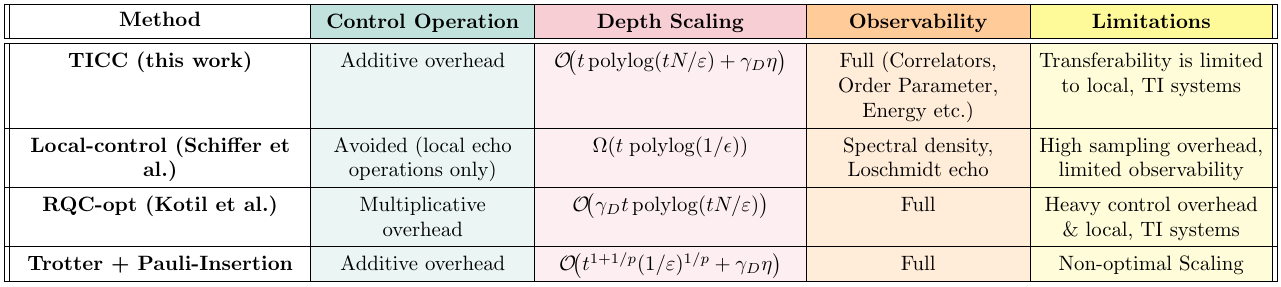}
    \caption{Conceptual comparison of different approaches for performing Quantum Phase Estimation. We compare our method (TICC) to: \textbf{(1)} Local control variant proposed by Shiffer et al. \cite{schiffer2025hardwareefficientquantumphaseestimation} \textbf{(2)} Controlling every gate in a compressed time evolution circuit (RQC-opt \cite{rqcopt1} taken as representative) and \textbf{(3)} Pauli string insertion into a Trotterized circuit. Scaling lower bound given for the local control variant has been formalized in Eq. \ref{eq:gate_complexity_lower_bound}. }
    \label{fig:tab2}
\end{table*}

In order to demonstrate the efficiency of our compression protocol, we use our compressed circuits to simulate a quantum spin system with antiferromagnetic coupling on a triangular lattice. We achieve relative ground state energy errors below 1\% for the transverse-field Ising model (TFIM) with antiferromagnetic coupling on a 4$\times$4 triangular lattice, using the noise-aware emulator of a commercially available quantum device.

\section{Background}
\label{sec:sec2}
In order to encode the time evolution operator of a generic Hamiltonian into a quantum circuit, there is a fundamental lower bound for the required number of queries to be made to an oracle, encoding the Hamiltonian. This lower bound scales at least linearly with evolution time $\Omega(t)$ \cite{Berry_2006, Gu_2021}.

Widely used, product formula-based approaches such as the Trotter-Suzuki decomposition \cite{Su91} scales considerably worse than this lower bound with 
\begin{equation}
    \mathcal{O}\left(t^{1+1/p} \left(1/\epsilon\right)^{1/p}\right)
\end{equation}
for the Trotter-Suzuki decomposition \cite{Childs_2021}, where $p$ is the so-called splitting order and $\epsilon$ is the error metric in terms of the spectral norm (defined in Eq. \ref{eq:spectral_norm}), which we will use throughout the rest of the paper. Although, by increasing the order $p$ arbitrarily high, one can get nearer to a linear scaling behavior (with respect to the evolution time), this brings an exponentially increasing overhead $\mathcal{O}( 5^{p/2-1} N)$ for $N$ being the system size of the Hamiltonian \cite{Wiebe_2010, Avtandilyan_2024}. Although their scaling with respect to evolution time and target accuracy are not optimal (for the generic, optimal bound please refer to our derivation in App. \ref{sec:A1}), these methods are often favored and employed with low splitting order ($p=1, 2$), due to their low implementation overhead.

One way to achieve near-optimal scaling with overheads comparable to those of first- or second-order Trotterization was proposed by Kotil et al. \cite{rqcopt1}, in the form of a variational optimization, targeting  translationally invariant (TI), local Hamiltonians. In this Riemannian Quantum Circuit Optimization (RQC-opt) protocol, one employs a set of two qubit unitaries as the parameter set $\{V_0, V_1, \dots\}$ of a cost function:
\begin{equation}
\label{eq:cost_1}
    f(V_0, V_1, \dots ) = -\text{Re}\{\text{Tr}\left[ U(t)^{\dagger} \hspace{0.1cm} W(V_0, V_1, \dots) \right]\}
\end{equation}
where $U(t) = e^{-i H t}$ is the full time evolution operator, acquired through matrix exponentiation and $W(V_0, V_1, \dots)$ is the Ansatz unitary that $\{V_0, V_1, \dots\}$ two qubit gates constitute, when laid out in a brickwall circuit, as visualized in Fig. \ref{fig:fig1}.a. Ref. \cite{rqcopt1} also formalizes how to compute gradients and Hessians of such a cost function. The proposed Ansatz construction, uses the same  unitary  $V_i$ for each  gate within a layer $i$ of the brickwall, such that the Ansatz is translationally symmetric. Authors optimize these gates $\{V_0, V_1, \dots\}$ such that their brickwall Ansatz $W(V_0, V_1, \dots)$ approximates the time evolution operator of a small system (e.g. 4 or 6 qubits). They then show that this optimized outcome $\{V^*_0, V^*_1, \dots\}$ can be used to approximate the time evolution operator of larger systems of the same, translationally-invariant Hamiltonian with negligible increase in approximation error (see Fig. 5.b of Ref. \cite{rqcopt1}). Hence, the translational invariance of the Ansatz circuit plays a crucial role for the transferability of optimized gates.

This small to large system transferability is ensured for a sufficiently small evolution time $t \leq t_{\text{max}}$ by the locality of the Hamiltonian, as the propagation speed of operator spreading and correlations generated by local interactions is finite \cite{LR}. This propagation speed, known as Lieb-Robinson velocity $v_{\text{LR}}$, governs the information propagation on a quantum lattice system with short range interactions. This helps us to set a maximal evolution time, for which one can employ an $N$-qubit brickwall Ansatz in the optimization and can reuse the optimized gates to simulate larger systems. As the maximal distance between two sites on an $N$-qubit lattice of $D$ dimensions, scales as $\mathcal{O}(N^{1/D})$, we can argue
\begin{equation}
\label{eq:tmax_scaling}
    t_{\text{max}} = \mathcal{O}\left( \frac{N^{1/D}}{v_{\text{LR}}}\right).
\end{equation}
Through numerical simulations, we validate this scaling of the maximal simulation time. We also show that circuit depth of the Riemannian Circuit Optimization achieves the near-optimal scaling 
\begin{equation}
\label{eq:rqc_scaling}
\mathcal{O}(t \text{ polylog}(t N/\epsilon)).
\end{equation}
This scaling behaviour was not formalized in Ref. \cite{rqcopt1}. In this work, we run numerical investigations on validating how RQC-opt satisfies this achievable scaling derived in Ref. \cite{Haah_2021} and also formalize that this bound is near-optimal in the $t \gg \log(1/\epsilon)$ limit.  For more details, we refer the reader to Appendix \ref{sec:A1}.

In order to control the time evolution, one naive approach could be directly controlling each of these optimized gates and decomposing the resulting unitaries into an ensemble of hardware native two qubit gates. However, this causes a significant, multiplicative control overhead as each controlled-two qubit gate would be synthesized into $\gamma_D \simeq 15$ native two qubit gates \cite{Barenco_1995, Shende_2004}. On the other hand, one can try to optimize the Ansatz circuit to approximate the controlled evolution directly. This, however, would break the translational symmetry of the Ansatz through introducing an ancilla and losing the small-to-large system transferability property.

An alternative approach can be used when the quantum protocol, in which one employs the controlled evolution, includes a single ancilla qubit \cite{Babbush_2018, qetu}. Then, the following equivalence holds:
\begin{multline}
\label{eq:cU_equivalence}
    \ketbra{0}_{\text{anc}} \otimes I + \ketbra{1}_{\text{anc}} \otimes e^{-i H t} \iff \\
    \ketbra{0}_{\text{anc}} \otimes e^{i H t/2} + \ketbra{1}_{\text{anc}} \otimes e^{-i H t/2}
\end{multline}
within each energy eigenspace of $H$, differing only by a global phase factor that depends on the eigenvalue. This suggests that controlling the evolution direction is equivalent to controlling the evolution, when we are interested in simulating only a single energy eigenspace, which is the case for ground state preparation or ground state energy estimation. We also stress that, this equivalence does not necessitate system qubits to start from an eigenstate of the Hamiltonian. In most algorithms with controlled time evolution, measuring the ancilla projects the system qubits onto an eigenmanifold; effectively justifying the equivalence in Eq. \ref{eq:cU_equivalence}.

It is clear how such an equivalence holds when there is only one ancilla qubit in the algorithm, which is the case for the Iterative Phase Estimation, whose circuit representation is given in Fig. \ref{fig:fig1}.b (top). Luckily, such an equivalence can also be shown for the standard, Quantum Fourier Transform (QFT) based QPE \cite{nielsen_chuang}. For details, please refer to App. \ref{sec:A4}.

Building upon the equivalence given in Eq. \ref{eq:cU_equivalence}, authors in Ref. \cite{qetu} propose to decompose the Hamiltonian into a sum of sub-Hamiltonians $\{H_i\}_{i=1}^{\eta}$, where $\eta$ denotes the total number of sub-Hamiltonians. These sub-Hamiltonians are chosen such that one can find easy-to-control unitaries $\{K_i\}_{i=1}^{\eta}$ (e.g. Pauli strings) for each sub-Hamiltonian $H_i$, such that they anti-commute:
\begin{equation}
    \label{eq:ham_decomposition}
    H = \sum_{i=1}^{\eta} H_i \hspace{0.2cm} ; \hspace{0.5cm} K_i^{\dagger} H_i K_i = -H_i.
\end{equation}
Then, inserting the controlled versions of these $\{K_i\}_{i=1}^\eta$ unitaries before and after each $e^{-iH_it/2}$ layer of an appropriate Trotterization of $e^{-iHt/2}$, is equivalent to controlling the time evolution. 

Despite the promisingly low overhead of this method, the unfavorable asymptotic scaling of product-formula based implementations limits the achievable evolution times and accuracies, as required gate counts quickly exceed the coherence limits of current quantum devices. 

Aimed at tackling this problem and inspired by the aforementioned Pauli string insertion idea, we develop a compression protocol that achieves the near-optimal asymptotical scaling of RQC-opt, and also reduces the multiplicative control overhead $\gamma_D$ to an additive factor.

\begin{figure}
    \centering
    \includegraphics[width=0.99\linewidth]{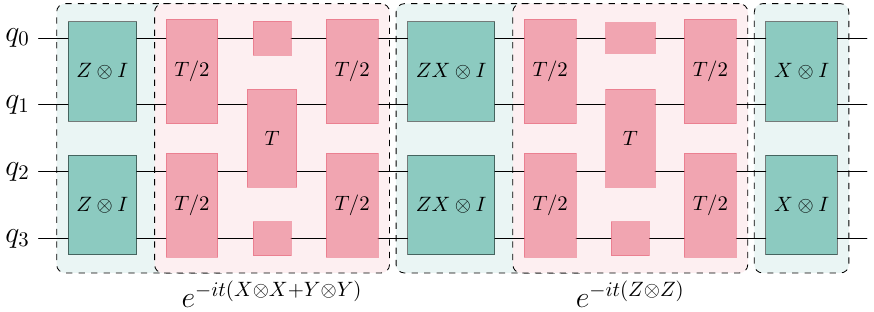}
    \caption{Ansatz employed to optimize a TICC circuit of time $T$, for one dimensional, isotropic Heisenberg model (\ref{eq:isotropic_HM}). Subscripts and gate labels indicate the starting point of the optimization, and they correspond to a splitting with respect to the decomposition $H_1 = \sum_{\langle i,j \rangle} X_iX_j+Y_iY_j$ and $H_2 = \sum_{\langle i,j \rangle} Z_iZ_j$. This decomposition implements a splitting of the form given in Eq. \ref{eq:ham_decomposition} with $\eta =2$, $K_1 = Z \otimes I \otimes Z \dots$ and $K_2 = X \otimes I \otimes X \dots$. Because each red block employs $\gamma = 3$ layers, we start the optimization from an initial point that corresponds to implementing each $e^{-iT H_i}$ term with a second order Trotterization, where the gate labels in red blocks indicate the time steps $t$ used in each $e^{-i t H_i}$ term. Although we start from a separable state for the green layers, the Ansatz allows entanglement between sites $\{(2j, 2j+1)\}_{j=0}^1$.}
    \label{fig:fig2}
\end{figure}

\section{Translationally Invariant Compressed Control (TICC)}
\label{sec:sec3}

Our proposal is based on modifying the cost function (\ref{eq:cost_1}) and Ansatz of the RQC-opt protocol, in a way that is inspired by the Pauli string insertion idea explained in Sec. \ref{sec:sec2}. In our Translationally Invariant Compressed Control (TICC) protocol, we introduce a structured partition of all variational parameters into two sets, where
\[
V = (V_0, V_1, \dots)
\]
is the full parameter set describing all layers of the variational circuit 
$W(V)$, and
\[
\Tilde{V} \subset V
\]
is a reduced parameter set.

In this optimization protocol, the cost function consists of two additive parts
\begin{multline}
\label{eq:cost_2}
f(V) = -\text{Re}\{\text{Tr}\left[ U(t/2)^{\dagger} \hspace{0.1cm} W(V) \right]\} \\ -\text{Re}\{\text{Tr}\left[ U(-t/2)^{\dagger} \hspace{0.1cm} W(\Tilde{V}) \right]\}.
\end{multline} The optimal gates that minimize such a cost function, constitute a quantum circuit that approximates:
\begin{equation}
     \ketbra{0}_{\text{anc}} \otimes U(-t/2) + \ketbra{1}_{\text{anc}} \otimes U(t/2)
\end{equation}
once we control the gates in $V \textbackslash \Tilde{V}$ with our ancilla and leave the gates in $\Tilde{V}$ layers uncontrolled. That way, ancilla qubit controls a switching behavior between $+t/2$ and $-t/2$ evolution, which is exactly what the circuit equivalence in Eq. \ref{eq:cU_equivalence} represents.

Such an Ansatz is desirable for multiple reasons. This way, one can lower the control overhead compared to controlling every layer with the ancilla, because with this construction we only have to control a sub-set of the gates, namely: $V \textbackslash \Tilde{V}$. This subset, which we will also refer to as \textit{control layers}, is visualized as the green layers of the circuit visualized in Fig. \ref{fig:fig2}. 

Inspired by the Pauli string insertion idea mentioned in Sec. \ref{sec:sec2}, we can pick the number of control layers ($|V\textbackslash \Tilde{V}|$) equal to $\eta+1$ according to a Hamiltonian decomposition that satisfies the constraints mentioned in Eq. \ref{eq:ham_decomposition}. Crucially, this number does not depend on the evolution time but only on the Hamiltonian structure. As a direct result of this, the control overhead factor $\gamma_D$ does not affect the whole circuit depth as a multiplicative factor but is an additive factor  in evolution time.

More formally, the number of gates in $\Tilde{V}$ (red layers in Fig. \ref{fig:fig2}) will have to scale like the aforementioned $\mathcal{O}(t\text{ polylog}(t N/\epsilon))$ (\ref{eq:rqc_scaling}) scaling of RQC-opt. Combining the decomposition factor $\gamma_D$ with the number of control layers $\eta+1$, we arrive at:
\begin{equation}
    \mathcal{O}\left(t \text{ polylog}(t N/\epsilon) +  \gamma_D (\eta+1) \right)
\end{equation}
for the number of circuit layers in the full Ansatz after the decomposition of control layers into arbitrary two qubit gates. In the large $t$ limit $t \gg \gamma_D (\eta+1)$, we recover the same scaling we give for RQC-opt from Eq. \ref{eq:rqc_scaling}.  Thereby our main contribution to this scaling behaviour is to turn the control overhead factor from being multiplicative to additive in time. In the $t \gg \log(1/\epsilon)$ limit, one can also argue that this scaling is near-optimal.  For a more formal treatment of this scaling bound, we refer the reader to App. \ref{sec:A1}.

An important thing to note is that the ancilla qubit is unknown to the optimizer. It is crucial for the translational invariance that the Ansatz circuit we run the optimization on, does not include an ancilla qubit, because otherwise the ancilla would break the translational symmetry. This symmetry is crucial to claim transferability of the optimized gates from small to larger systems. Only after the optimization yields the optimal gates $V^*$ and once we want to employ these optimized gates to encode the controlled evolution, we synthesize controlled versions of the gates in $V^* \textbackslash \tilde{V}^*$.

\begin{table*}
    \centering
    \includegraphics[width=0.99\linewidth]{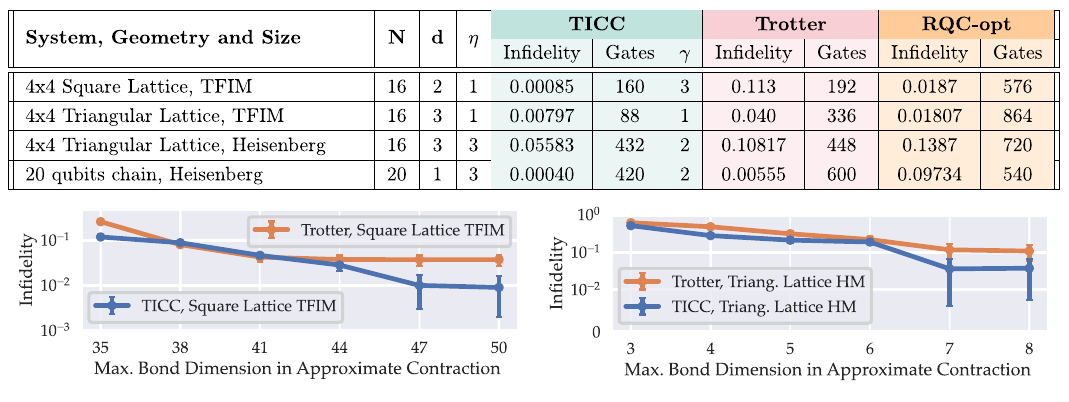}
    \caption{ \textbf{(Top)} Evolution infidelity (\ref{eq:ev_infidelity}) vs. the total number of arbitrary two qubit gates used to encode the controlled time evolution of different systems on square and triangular lattices (2D TFIM with transverse field strength $g=1.5$ (\ref{eq:TFIM}) and 2D Heisenberg (\ref{eq:ansisotropic_HM})), as well as a linear chain of 20 qubits (1D Heisenberg model (\ref{eq:ansisotropic_HM})), for evolution time $t = 0.125$. Trotter method is based on using the idea of inserting controlled Pauli strings (explained in Sec. \ref{sec:sec2}) with basis time steps $\Delta t = 0.125$ for 4$\times$4 square lattice TFIM and $\Delta t = 0.0625$ for the rest. For the 1D Heisenberg model, TICC and RQC-opt gates were optimized on a 6 qubits chain (and reused on the 20 qubits system for evaluation). For all evolution infidelity calculations, we evolve 20 randomized states with each of the different methods and average the infidelity (resulting standard deviation of the averaging is below 0.1\% for each calculation). All system geometries assume periodic boundary conditions. Controlled two qubit gates (employed in RQC-opt and in control layers of TICC) are decomposed into $2\text{-}9$ arbitrary two-qubit gates, using an approximate decomposer with target spectral norm error $\epsilon_D \leq 10^{-5}$ per decomposition.  For each system we report the qubit number $N$, number of inequivalent nearest-neighbor permutations of the geometry $d$, the number of uncontrolled layers between each control layer of TICC $\gamma$, as well as $\eta$ for given Hamiltonians (see Eq. \ref{eq:ham_decomposition}). \textbf{(Bottom)} For the transferability of the optimized gates to larger systems, we run a sanity-check by investigating the evolution infidelity for TFIM \eqref{eq:TFIM} on 6$\times$6 square lattice and HM \eqref{eq:ansisotropic_HM} on 6$\times$6 triangular lattice, through PEPS simulations of bond dimension 4. PEPS contraction is performed approximately with the \texttt{auto-hq} (for square lattice) and \texttt{hyper-compressed} (for triangular lattice) methods of \texttt{QUIMB}. We average evolution infidelities of 10 random initial states with bond dimension 1, after evolving them approximately (with TICC vs. Trotter) and computing their overlap with states evolved with finer Trotterization (fourth order) as reference. }
    \label{fig:fig3}
\end{table*}

We observe that the choice of Ansatz structure and initial point of the optimization is crucial to the convergence performance of our protocol. We choose an Ansatz structure that corresponds to a Hamiltonian decomposition of a form given in Eq. \ref{eq:ham_decomposition}. So the layers in $V\textbackslash \Tilde{V}$ subset are in analogy to the anti-commuting Pauli string layers $K_i$. As the starting point for the optimization, we choose a Trotterization of such a decomposition for the $\Tilde{V}$ layers and choose the corresponding $K_i$ strings for the initialization of $V\textbackslash \Tilde{V}$ layers. The specifics of this Trotterization (i.e. splitting order, time steps in each layer etc.) depends on the number of layers chosen for the TICC Ansatz circuit.

As a concrete example, we visualize in Fig. \ref{fig:fig2}, an Ansatz on four sites; employed for approximating the time evolution of one dimensional, isotropic Heisenberg model, defined as:
\begin{equation}
\label{eq:isotropic_HM}
    H_{\text{iHM}} = \sum_{O \in \{X, Y, Z\}}\sum_{\langle i, j \rangle} O_i O_j
\end{equation}
where $X_i, Y_i, Z_i$ are Pauli X, Y, Z operators acting on site $i$ and $\langle i, j\rangle$ sums over nearest neighbors. 

For encoding two dimensional Hamiltonians, the decomposition of the form given in Eq. \ref{eq:ham_decomposition} becomes more involved. In this case, one has to decompose not only with respect to the Pauli terms but also with respect to vertical and horizontal permutations of nearest neighbors. We give further details of our two dimensional Ansatz circuits in App. \ref{sec:A2}.

In Table \ref{fig:tab2}, we compare our proposed method to the other, state of the art proposals for making QPE more hardware efficient, by listing the limitations and advantages of each considered method.

\section{Results}
\label{sec:sec4}
We benchmark our compression protocol through numerical simulations by comparing the evolution infidelity, defined as

\begin{equation}
\label{eq:ev_infidelity}
    \epsilon_{\text{ev}} = 1-\mathbb{E} \left[ | \langle v | U(t)^{\dagger} \Tilde{U} | v\rangle |^2 \right]_{\mathcal{H}}
\end{equation}
where $U(t)$ is the exact time evolution unitary, $\Tilde{U}$ is the approximate unitary (TICC, Trotterization, RQC-opt etc.) and $v$ is sampled from the uniform distribution, induced by the Haar measure $\mathcal{H}$.  These states are $N$+1 qubit states, for $N$ being the system size (and plus one for ancilla). We report optimization results, run for two different quantum spin systems: Antiferromagnetic, transverse field Ising model (TFIM), defined as
\begin{equation}
\label{eq:TFIM}
    H_{\text{TFIM}} = \sum_{\langle i, j \rangle} Z_iZ_j + g \sum_{i=0}^{N-1} X_i
\end{equation}
where $g$ is the transverse field strength, and Heisenberg model in a field (HM)
\begin{equation}
\label{eq:ansisotropic_HM}
    H_{\text{HM}} = \sum_{O \in \{X, Y, Z\}} \left( \sum_{\langle i, j \rangle} O_iO_j + h_O \sum_{i=0}^{N-1} O_i \right)
\end{equation}
with the field parameters set to $(h_X, h_Y, h_Z) = (3, -1, 1)$. We investigate these two systems on different one and two dimensional geometries with periodic boundaries. Particularly, we run optimizations on a 4$\times$4 square lattice, where each site has four nearest neighbors separated at 90° from each other, as well as a 4$\times$4 triangular lattice where each site has six nearest neighbors separated at 60°. Such a triangular connectivity is particularly challenging to simulate due to two reasons: Higher connectivity per qubit causes higher gate counts and triangular connectivity introduces geometric frustration for antiferromagnetic interactions.

\begin{figure*}
    \centering
    \includegraphics[width=0.9\linewidth]{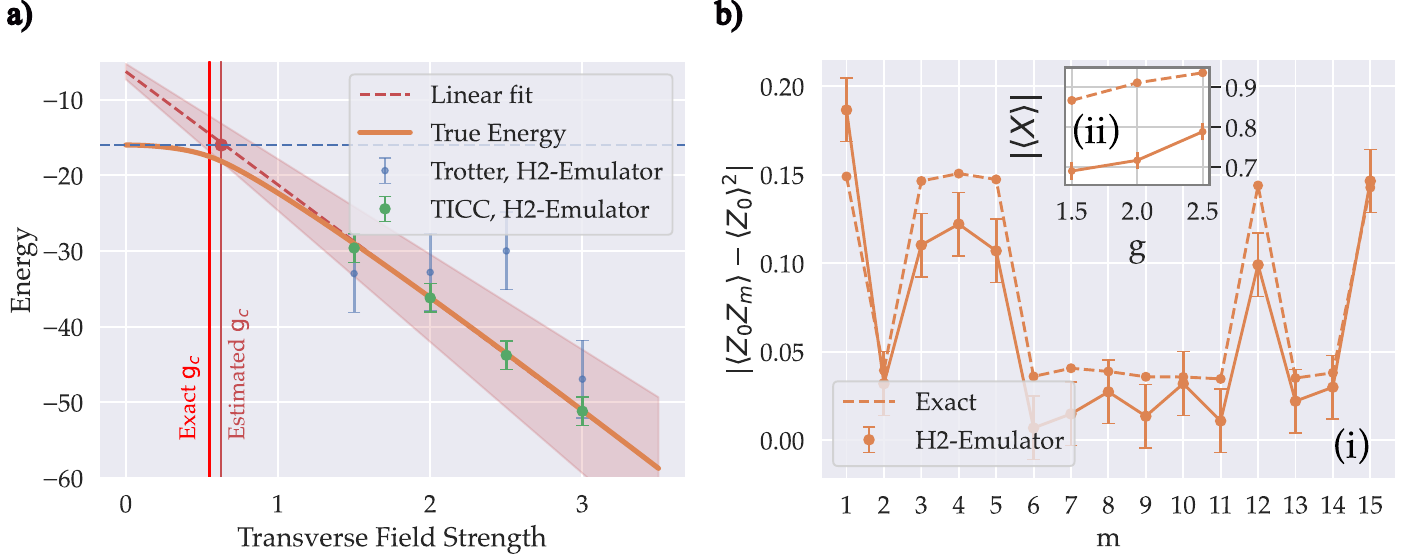}
    \caption{\textbf{(a)} Iterative QPE, employed for ground state energy estimations of the antiferromagnetic TFIM (\ref{eq:TFIM}) with various field strengths on a triangular 4×4 lattice. We start from the initial energy estimates at \{-40, -45, -50, -60\} for the field strengths considered $g \in \{1.5, 2, 2.5, 3\}$. These values were interpolated from the ground state energy solutions of 2$\times$2 and 3$\times$3 system sizes (which were solved by exact diagonalization) and then rounded up to the nearest multiple of 5. We run the iterative scheme (whose details are explained in App. \ref{sec:A_QPE}) for times $t=0.125$ and $t=0.25$ ($k \in \{-3, -2\} $). To encode the controlled time evolution operator, we employ two approaches: Second order Trotterization with Pauli string insertion, explained in Sec. \ref{sec:sec2} (blue) and TICC (green). TICC circuits employ 184 hardware-native ZZPhase gates per phase estimation circuit, whereas Trotter circuits employ 336. We use 200 shots per expectation value measurement for both methods. With all of the field strengths considered, TICC delivers sub-1\% relative energy error (\{0.7\%, 0.1\%, 0.2\%, 0.2\%\} for the given transverse field strengths) with $\pm 1.5\%$ variation. These error bars are calculated through the protocol explained in App. \ref{sec:A_QPE} and visualized in Fig.\ref{fig:fig1}.c, with  amplification factors in the range $(0.72, \, 1.8)$ and $1.62 \pm 0.1$ as average,  for the TICC results.  We interpolate between the asymptotic linear behavior of the energy at large field strengths and the classical energy (at zero field) to qualitatively estimate the pseudo–phase transition point ($g_c$). \textbf{(b)} Correlation function at transverse field $g=2$ \textbf{(i)} and (dis)order parameters $|\langle X\rangle|$ at $g\in\{1.5, 2, 2.5\}$ \textbf{(ii)}, sampled through QFT based QPE protocol for basis time $t_0=0.1$ with two ancilla qubits. Controlled time evolutions are implemented with TICC circuits. We use 2000 shots for expectation value measurement per each basis (X and Z) and the error bars represent the statistical, standard deviation. Each circuit employs 366 hardware native ZZPhase gates for the full, QFT appended protocol. Results in both plots are obtained from simulations on the noise-aware emulator of the Quantinuum H2 device, whose noise parameters are listed in App. \ref{sec:A3}.}
    \label{fig:fig4}
\end{figure*}

In Table \ref{fig:fig3}, we visualize our numerical benchmarks of comparing TICC to 1) Pauli string insertion based second order Trotterization 2) directly controlling RQC-opt gates. Moreover, we run minimal PEPS simulations with open source python library $\textsc{quimb}$, to verify the transferability of the 4$\times$4 results to a 6$\times$6 lattice systems. 

For the considered geometries, we group nearest-neighbor permutations into equivalence classes under lattice translations, where permutations that differ only by a cyclic shift of the sites are considered equivalent. This grouping is necessary for implementing the decomposition given in Eq. \ref{eq:ham_decomposition} for two dimensional systems, as permutations belonging to different equivalence classes cannot be controlled with the same unitary $K_i$. Hence, one has to decompose the Hamiltonian in terms of these permutation groups, in addition to a decomposition in terms of the Pauli terms. For each lattice geometry, we report the number of inequivalent nearest-neighbor permutations $d$ as follows: for the one-dimensional chain, two permutations collapse into a $d=1$ equivalence class; for the square lattice, four permutations reduce to $d=2$ distinct classes; and for the triangular lattice, six permutations reduce to $d=3$ classes. We explicitly give these permutations for each inequivalent permutation group in App. \ref{sec:A2}.

We give the Ansatz construction parameter $\eta$ (\ref{eq:ham_decomposition}) for each of the considered systems per inequivalent nearest-neighbor permutation as follows: $\eta=1$ TFIM (\ref{eq:TFIM}) and $\eta=3$ for the Heisenberg model (\ref{eq:ansisotropic_HM}). For TFIM, one Pauli string:
\begin{equation}
    K = Y \otimes Z \otimes Y \otimes Z \otimes \dots
\end{equation}
already anti-commutes with the Hamiltonian in each inequivalent permutation group (hence $\eta=1$). For the HM, here is an example decomposition with corresponding Pauli strings $K_i$, that we use also for the initialization of the TICC optimization protocol of this system
\begin{equation}
    H^{(1)}_{\text{{HM}}} = \sum_{\langle i, j \rangle} X_iX_j + h_Y \sum_{i=0}^{N-1} Y_i \hspace{0.1cm},  K_1 = \bigotimes_{i=1}^{N/2}  (X \otimes Z),
\end{equation}
    
\begin{equation}
    H^{(2)}_{\text{{HM}}} = \sum_{\langle i, j \rangle} Y_iY_j + h_Z \sum_{i=0}^{N-1} Z_i \hspace{0.1cm}, K_2 = \bigotimes_{i=1}^{N/2}  (X \otimes Y),
\end{equation}
\begin{equation}
    H^{(3)}_{\text{{HM}}} = \sum_{\langle i, j \rangle} Z_iZ_j + h_X \sum_{i=0}^{N-1} X_i \hspace{0.1cm}, K_3 = \bigotimes_{i=1}^{N/2}  (Y \otimes Z).
\end{equation}

These parameters $(d, \eta)$ are determined by the Hamiltonian structure and geometry but not scaled with targeted evolution time. Using these parameters, we can formalize the total two qubit gate count
\begin{equation}
\label{eq:total_gate_formula}
    \frac{N}{2} d (\gamma \hspace{0.05cm} \eta + \gamma_D (\eta+1))  
\end{equation}
where $N$ is the number of sites in the considered system size and $\gamma_D$ is the decomposition overhead factor of the controlled two qubit gates, i.e., the number of elementary two-qubit gates required to implement a single controlled two-qubit unitary. For the numerical benchmarking in Table \ref{fig:fig3}, we apply this decomposition with the open source, approximate decomposer \texttt{UniversalQCompiler} \cite{UniversalQCompiler2020}. We set the target spectral norm error of the decomposition to $\epsilon_D \leq 10^{-5}$, and observe that this corresponds to $\gamma_D \in [2, 9]$ depending on the specific two-qubit unitary to control.

Previously introduced trace norm cost (used in Eq. \ref{eq:cost_2}) and evolution infidelity (\ref{eq:ev_infidelity}) are known to be equal up to an insignificant constant $\frac{2^{N+1}}{2^{N+1}+1}$  \cite{Nielsen_2002, Khatri_2019}. We can use this equivalence to obtain an error budget relation for the evolution infidelity
\begin{equation}
        \epsilon_{\text{ev}} \leq \epsilon_{\rightarrow}  + \epsilon_{\leftarrow} + \frac{N}{2} d \gamma_D (\eta+1) \epsilon_D
\end{equation}
where $\epsilon_{\rightarrow}$ is the forwards time evolution approximation error (first term in Eq. \ref{eq:cost_2}), $\epsilon_{\leftarrow}$ is the backwards time evolution approximation error (second term in Eq. \ref{eq:cost_2}), $\frac{N}{2} d  \gamma_D  (\eta+1)$ is the number of controlled two qubit gates, decomposed into hardware native gates, with an approximation error $\lesssim \epsilon_D$. For the derivation of this bound, we used the triangle inequality and a standard telescoping sum argument \cite{tan2025unitarysynthesisfewert}. Moreover, one can typically treat $\gamma_D = \mathcal{O}(\text{polylog}(1/\epsilon_D))$ according to standard approximate decomposition schemes \cite{dawson2005solovaykitaevalgorithm}.

We also employ our compressed circuits in phase estimation protocols using the noise aware emulator of Quantinuum H2 device to demonstrate that TICC provides such considerably low gate counts for the systems considered in this work, that currently available quantum hardware can deliver qualitatively meaningful results. More concretely, our emulation protocols for iterative QPE use 184 hardware native ZZPhase gates for the 4$\times$4 triangular lattice system. These ZZPhase gates are native to most ion trap platforms and implement
\begin{equation}
    \text{exp}(-i\frac{\pi \Theta}{2} Z \otimes Z)
\end{equation}
with gate parameter $\Theta$. For a potential up-scaled demonstration, this would correspond to 414 ZZPhase gates for the 6$\times$6 system (as gate count is linear in system size, see Eq. \ref{eq:total_gate_formula}). Results of these QPE protocols are visualized in Fig. \ref{fig:fig4} and details of the used protocols are reported in App. \ref{sec:A_QPE}.

\section{Discussion}
\label{sec:sec5}
As reported in previous sections, our compression proposal can reduce the number of gates for various classes of Hamiltonians. However, in all of these classes, translational invariance and locality of interactions are explicitly assumed. Our proposed method is not directly applicable to systems with disorder (breaking translational symmetry) and non local interactions. 

Although it is not enforced formally, the existence of a decomposition scheme according to Eq. \ref{eq:ham_decomposition} is practically important. This is due to the fact that in all our optimization results, the Ansatz is constructed with a form inspired by this decomposition and also Trotterization of such a decomposition is chosen as the starting point of the optimization. This, however, does not necessarily restrict the classes of Hamiltonians one can apply TICC to, as any Hermitian operator admits an exact expansion in the Pauli basis \cite{nielsen_chuang}. For each non-identity Pauli term (or suitable groupings thereof), one can always find a Pauli string that anti-commutes with it. The identity component only contributes to a global energy shift and is physically irrelevant.  The number of such Pauli blocks $\eta$ does not depend on the targeted evolution time, hence the control overhead is additive in time. Importantly, for highly non-local systems this number can grow with $N$ exponentially, in the worst case. However, for most of the geometry classes of interest where interactions are $k$-local and chosen from a fixed finite set of local interaction types (as is typical in lattice models \cite{Childs_2019}), $\eta$ is fixed by $k$ and does not depend on $N$.

A direct and unavoidable limitation of TICC is the limit in the maximal evolution time $t_{\text{max}} = \mathcal{O}(N^{1/D})$ (\ref{eq:tmax_scaling}) for the given Ansatz size $N$ (and dimensionality $D$), in order for the "small to large system transferability" to hold. If the targeted evolution time $t$ exceeds the maximal evolution time of the given Ansatz size, one can optimize TICC gates for a time step of $\Delta t < t_{\text{max}}$ and linearly repeat the block $t/\Delta t$ many times. This, however, would expectedly make the approximation error linearly accumulate. Alternatively, one can increase the Ansatz size $N$, if computational resources allow. If computational resources do not allow for optimizing for the Ansatz system sizes needed, one can find Trotterization based implementation still more favorable.  

Another important point is the state preparation aspect of the QPE. Results reported in this work focused on a regime, relatively far away from the (pseudo) phase transition point. In this regime, it is very low cost to approximate the ground state with high enough overlap (above $\sim$53\% for the purposes of iterative QPE). However, around criticality the ground state preparation continues to be a major bottleneck for Hamiltonian simulation. This is where the versatility of TICC comes into play, as many optimally scaling ground state preparation algorithms (e.g. eigenstate filtering \cite{qetu, aff}) rely on controlled time evolution and employing TICC in such protocols can help with state preparation costs.

We also note recent proposals that avoid global control of time evolution altogether for the QPE \cite{schiffer2025hardwareefficientquantumphaseestimation, Kanno2025TensorQPDE} but point out that such protocols are limited by high sampling overheads or reduced observability of the prepared quantum state. Here, we focus on making controlled time evolution itself more efficient, as it remains a central primitive beyond QPE.

\section{Conclusion and Outlook}
We have introduced the Translationally Invariant Compressed Control (TICC) protocol, encoding the controlled time evolution of translationally symmetric, local systems. We show that our method reduces large overheads of controlled two qubit gates and we argue that it scales near optimally with respect to evolution time and accuracy. Gate counts achieved by TICC in our QPE benchmarks indicate that our method can bridge the gap between early fault-tolerant and currently executable regimes, enabling demonstrations on systems as large as 6×6 for non-trivial Hamiltonians. Our emulations show that the powerful iterative QPE protocol can achieve sub-percent errors on current devices, when combined with the hardware efficiency of the TICC protocol.

Such algorithmic advances that lower the practical overhead of canonical quantum algorithms, bring us closer to demonstrate quantum advantage in the near future. Our focus in the upcoming projects is to integrate our compression scheme with a hardware capability, whose native two qubit gate operation has access to more than just one of the KAK-parameters, which can implement an arbitrary SU(4) gate more efficiently than a ZZPhase or CNOT decomposition scheme. Integrating this compression protocol with, e.g. a hardware native B-gate capability \cite{Zhang_2004, Wei_2024}, would automatically cut down the number of hardware native operations by 1/3. Such a powerful, closer-to-hardware compression can enable running QPE on e.g. 6$\times$6 triangular lattice TFIM with as few as 276 hardware native gates.

Moreover, a direct follow up to this work will focus on analytically formalizing the conditions under which we can expect guaranteed convergence from the optimization protocol. This would enhance the rigor of our scaling bounds. Similarly, we will further verify the transferability of two dimensional circuits with better computational resources (PEPS/MPS simulations with higher bond dimensions) and investigate the performance of TICC for more diverse classes of Hamiltonians (e.g. fermionic systems, lattice systems with increased connectivity).

\section*{Acknowledgments}
The author thanks Michael Lubasch, Matteo D'Anna, Christian Mendl, Isabel Le, Ivan Rojkov and Juan Carasquilla for insightful discussions during the conceptual development and numerical investigations of the project. The author also acknowledges the coastal town of Datça for providing a setting that was both inspiring and conducive to this work.

\appendix
\renewcommand{\thefigure}{A\arabic{figure}}
\setcounter{figure}{0}
\renewcommand{\thetable}{A\arabic{table}}
\setcounter{table}{0}
\renewcommand{\theHfigure}{A\arabic{figure}}
\renewcommand{\theHtable}{A\arabic{figure}}

\section{Asymptotic scaling of Riemannian Circuit Optimization}
\label{sec:A1}
As we argue in Sec. \ref{sec:sec3}, our TICC protocol achieves the same asymptotic scaling of RQC-opt, which we conjecture to be near-optimal in Sec. \ref{sec:sec2}. In this section, we show that there exists such a translationally invariant, brickwall circuit $W$; which approximates the time evolution unitary $e^{-iHt}$ with error

\begin{equation}
\label{eq:spectral_norm}
    \| W - e^{-iHt} \| \leq \epsilon 
\end{equation}
where $\|.\|$ is the spectral matrix norm and whose depth scales as given in Eq. \ref{eq:rqc_scaling}. Then, we demonstrate the convergence behavior of RQC-opt to such an optimum, through numerical simulations. 
\begin{theorem}[Existence of local approximants {\cite{Haah_2021}}]
\label{thm:existence}
For any 2-local lattice Hamiltonian $H$ on a lattice $\Lambda$ of dimension $D$ and $L^D=N$ qubits, any simulation time $t>0$, 
and accuracy parameter $0<\epsilon<1$, there exists a 2-local circuit $W$ with total gate count
\[
\mathcal{O}\hspace{0.1cm}\!\big(\frac{N}{2}t \,\mathrm{polylog}(N t / \epsilon)\big)
\]
and depth 
\[
\mathcal{O}\hspace{0.1cm}\!\big(t \,\mathrm{polylog}(N t / \epsilon)\big)
\]
such that $\| e^{-iHt} - W \| \le \epsilon$, assuming parallelization of $N/2$ gates per circuit layer.
\end{theorem}

\begin{figure}
    \centering
    \includegraphics[width=0.99\linewidth]{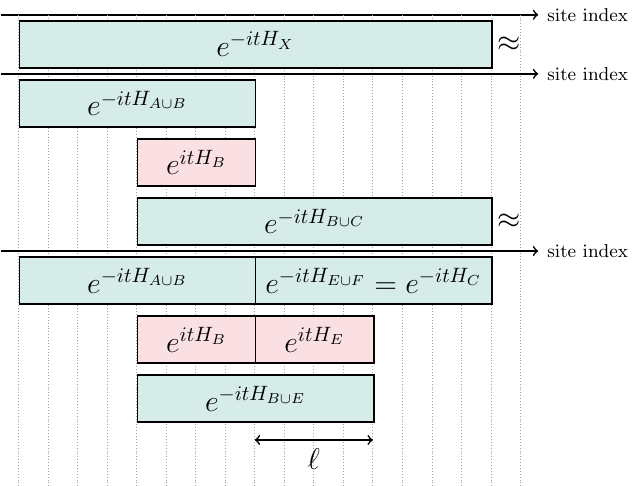}
    \caption{Two staged decomposition of the large unitary into smaller sub-unitaries, as given in Eqs. \ref{eq:decomposition} \& \ref{eq:decomposition2}. For the case of translationally invariant $H_X$ under local translations: $H_{A \cup B} = H_{C}$ and $H_B = H_E$, hence the resulting circuit is translationally symmetric. This decomposition is the same as the sequence used in \cite{Haah_2021}, with the assumption of $\norm{A \cup B} = \norm{C}$. }
    \label{fig:fig8}
\end{figure}

The authors of Ref. \cite{Haah_2021}, rely on the fundamental Lieb-Robinson bounds for the information propagation in local quantum systems \cite{LR, Hastings_2004, Hastings_2006, hastings2010localityquantumsystems}, to prove this bound and argue its near-optimality. There, authors use an iterative decomposition process of a large unitary $e^{-i H_X t_0}$ into:
\begin{multline}
\label{eq:decomposition}
    e^{-i H_X t_0} = e^{-i H_{A \cup B} t_0} e^{i H_{B} t_0} e^{-i H_{B \cup C} t_0} \\+ \mathcal{O}(\norm{H_B} \hspace{0.2cm} e^{-l(A, C)/\xi})
\end{multline}
where $t_0 = \mathcal{O}(1)$, $A \cup B \cup C = X$, $\{A, B, C\}$ are disjoint subsets of $X$, $l(A, C)$ is the lattice distance between $(A, C)$, and $\xi$ is a characteristic correlation length. In this notation, $H_X$ represents the Hamiltonian terms acting on the lattice subset $X$. This is Lemma 6 of \cite{Haah_2021}, which is proven by using the Lieb-Robinson theorem \cite{LR}. 

Then, one continues to decompose the $e^{-i H_{B \cup C} t_0}$ term (from Eq. \ref{eq:decomposition}) into:
\begin{equation}
\label{eq:decomposition2}
    e^{-i H_{B \cup C} t_0} \approx e^{-i H_{B \cup E} t_0} e^{i H_{E} t_0} e^{-i H_{E \cup F} t_0}
\end{equation}
where $\norm{B} = \norm{E}$, $E \cup F = C$ and $\{E, F\}$ are disjoint subsets of $C$. Visualization of this two step decomposition is given in Fig. \ref{fig:fig8}. Total approximation error of this two staged decomposition has the same upper bound as given in Eq. \ref{eq:decomposition}.

Authors of \cite{Haah_2021} show that starting from $H_X$ = $H$ (full Hamiltonian), choosing $\norm{B} = \ell \ll L$, iteratively continuing this decomposition (for each dimension) until there are only $(2 \ell)^D$-qubit gates left, decomposing these gates into two qubit gates using any known, optimal decomposer (e.g. \cite{Low2019hamiltonian, Berry_2015} - which brings the additional $\text{polylog}$ factors) and finally repeating the circuit $\mathcal{O}(t/t_0)$ many times for an arbitrary $t\geq t_0>0$ yields the scaling given in Theorem \ref{thm:existence}. For the full proof, please refer to \cite{Haah_2021}.

\begin{theorem}[Existence of translationally invariant, local approximants]
\label{thm:existence2}
For any 2-local, translationally invariant (TI) lattice Hamiltonian $H$ on a lattice $\Lambda$ of dimension $D$ and $L^{D}=N$ qubits ($L = 2^n$, $n \in \mathbb{N}$), any simulation time $t>0$, 
and accuracy parameter $0<\epsilon<1$; there exists a translationally invariant, 2-local circuit $W$ with depth 
\[
\mathcal{O}\hspace{0.1cm}\!\big(t \,\mathrm{polylog}(N t / \epsilon)\big)
\]
such that $\| e^{-iHt} - W \| \le \epsilon$, assuming parallelization of $N/2$ gates per circuit layer.
\end{theorem}

Proof of this Theorem can be shown by using the decomposition structure employed in the proof of Theorem \ref{thm:existence}. We first provide the proof for $D=1$ and later, argue a generalization to $D>1$. 

We show that the decomposition of $e^{-iHt_0}$ into $2\ell$-unitaries through the decomposition sequence given in Eqs. \ref{eq:decomposition} \& \ref{eq:decomposition2} (visualized in Fig. \ref{fig:fig8}) results in a TI circuit, for a TI Hamiltonian. To this end, we give the following sub-lattice definitions:
\begin{equation}
    \Lambda_j^{(k)} \coloneqq \{(j-1)2^{n-k}, \dots, j2^{n-k}-1\},
\end{equation}
\begin{equation}
    \Lambda_{\text{bd}(j)}^{(k)} \coloneqq \{j 2^{n-k}-\ell, \dots, j2^{n-k}+\ell-1\},
\end{equation}
\begin{equation}
    \Lambda_{\text{bd, L}(j)}^{(k)} \coloneqq \{j 2^{n-k}-\ell, \dots, j2^{n-k}-1\},
\end{equation}
\begin{equation}
    \Lambda_{\text{bd, R}(j)}^{(k)} \coloneqq \{j 2^{n-k}, \dots, j2^{n-k}+\ell-1\}.
\end{equation}
In this notation, each index represents a lattice point on this one dimensional geometry. We denote the resulting quantum circuit at each stage $k$ of the decomposition as $W(k)$ and define:

\begin{equation}
\label{eq:recurs_1}
    W(k) = \left\{\prod_{j=1}^{2^k} \text{exp}\left(-i H_{\Lambda_j^{(k)}}t_0\right) \right \} W_T(k)
\end{equation}
\begin{multline}
\label{eq:recurs_2}
W_T(k) \coloneqq \prod_{j=1}^{2^{k-1}} \text{exp}\left(i H_{\Lambda_{\text{bd, R}(j)}^{(k)} }t_0\right) \text{exp}\left(i H_{\Lambda_{\text{bd, L}(j)}^{(k)} }t_0 \right) \\ \text{exp}\left(-i H_{\Lambda_{\text{bd}(j)}^{(k)}}t_0\right)  W_T(k-1)
\end{multline}

\begin{multline}
    W_T(1) = \text{exp}\left(i H_{\Lambda_{\text{bd, R}(1)}^{(1)} }t_0\right) \text{exp}\left(i H_{\Lambda_{\text{bd, L}(1)}^{(1)} }t_0 \right) \\ \text{exp}\left(-i H_{\Lambda_{\text{bd}(1)}^{(1)}}t_0\right).
\end{multline}
In Eq. \ref{eq:recurs_1}, the index $j$ runs over all of the partitioned sub-lattices at step $k$. Each step partitions each sub-lattice into two, hence at step $k$ there are $2^k$ sub-lattices. In Eq. \ref{eq:recurs_2}, index $j$ runs over all the partition locations of step $k$ where each new partition brings three new terms ($\Lambda_{\text{bd, L}(j)}^{(k)} $, $\Lambda_{\text{bd, R}(j)}^{(k)}$ and $\Lambda_{\text{bd}(j)}^{(k)}$, corresponding to sub-lattices $B$, $E$ and $B \cup E$ from Fig. \ref{fig:fig8}). Number of new partitions at step $k$ equals to the number of sub-lattices from step $k-1$, hence $j$ runs from 1 to $2^{k-1}$. 

From these recursive definitions we already obtain an inductive proof for the Theorem \ref{thm:existence2} for $D=1$, as $W_T(1)$ (and hence $W(1)$) is TI and assuming $W_T(k-1)$ to be TI directly indicates $W(k)$ to be TI (induction step). 

For the generalization into $D>1$, we use the same roadmap, provided in \cite{Haah_2021}. There, authors consider iterating over the aforementioned sequence for each dimension $u \in (1, \dots, D)$, where each $W^u(k)$ circuit is extended along dimensions $(u+1, \dots, D)$. As this is a direct, translational extension, this step preserves translational invariance of the circuit, effectively extending our proof to $D>1$ circuits. This also means that, at the end of the iterative decomposition scheme at dimension $u$, we are left with unitaries of dimension $L^{D-u} \cross (2\ell)^u$. Once we iterate over all dimensions, the circuit is made up of only $(2\ell)^D$ qubit gates.

Existence of such circuits, however, does not guarantee a generic method to enable efficient convergence to such optima. In fact, for the non-convex landscapes, such guarantees are often out of reach. In this work, we report our empirical observation that the optimization protocol employed in RQC-opt, achieves efficient convergence to the global optimum, when the initialization is chosen to be close to the exact unitary. In practice, such an initialization is achieved through product formula based implementation of the time evolution operator whose form (i.e. splitting order) is determined by the number of layers in the Ansatz.

As the optimizer, we use a Riemannian Trust Region (RTR) solver \cite{Absil2008}, as proposed by \cite{rqcopt1}. The RTR parameters were set to the following values: acceptance threshold $\rho=0.125$, initial trust-region radius $\Delta_0 = 0.01$ and maximum radius $\Hat{\Delta}=0.1$. The trust-region subproblem is solved using the default truncated conjugate-gradient scheme. Optimization terminates once the variance of the loss function across successive iterations falls below the fixed threshold of $0.001$, indicating convergence.

\begin{figure}
    \centering
    \includegraphics[width=0.99\linewidth]{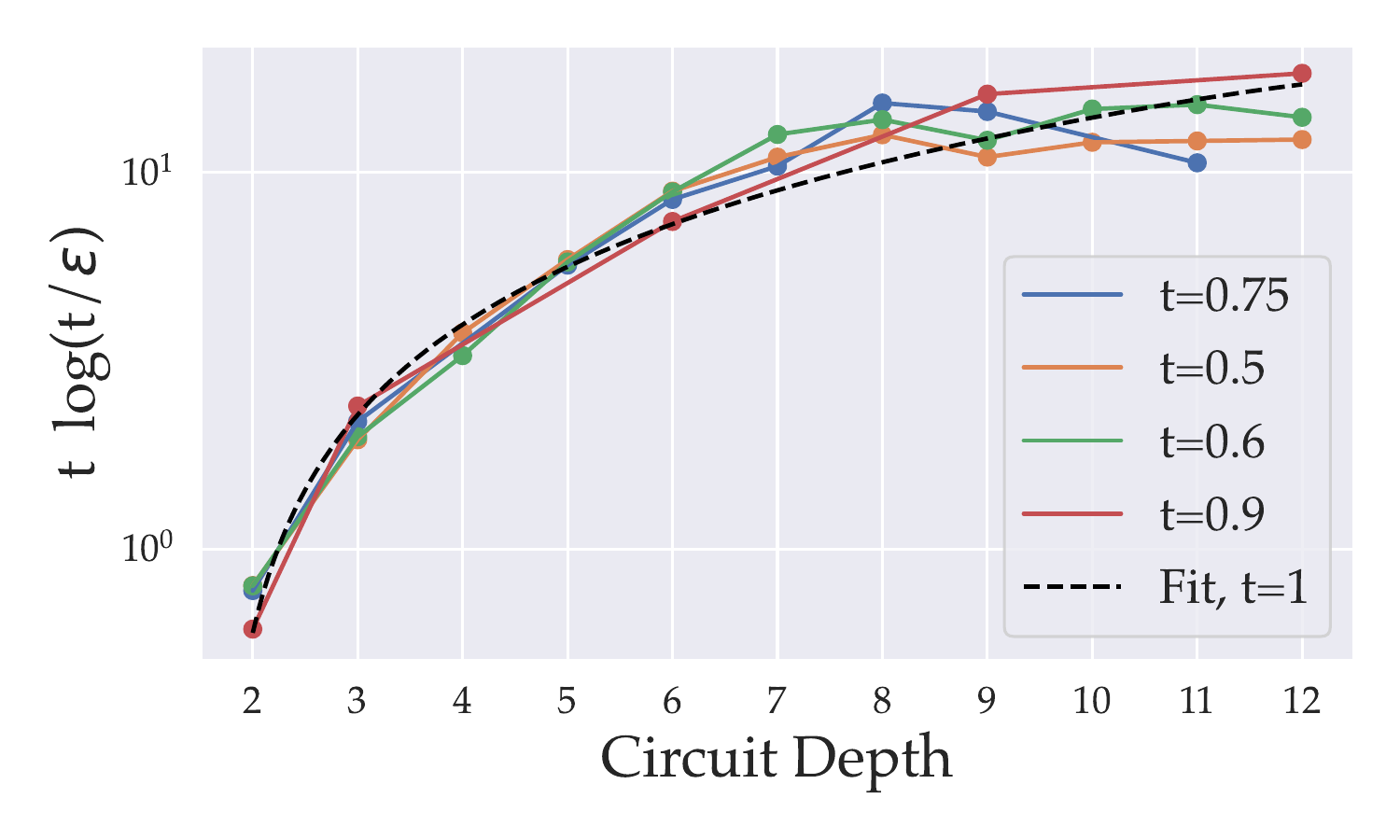}
    \caption{Numerical simulations of circuit depth scaling of the RQC-opt protocol. $\epsilon$ is the spectral norm distance error of the approximate time evolution unitary and $t$ is the evolution time. We fit the data points to the function given in Eq. \ref{eq:fit} and report that the optimal parameter corresponding to $c_1$ is approximately 1, suggesting that RQC-opt satisfies the near-optimal scaling $\mathcal{O}(t \log(t/\epsilon))$ for the circuit depth. }
    \label{fig:rqc_opt_numerical}
\end{figure}

We support our conjecture of efficient convergence, through numerical simulations aimed at investigating the circuit depth scaling of RQC-opt with respect to approximation error and evolution time. We run RQC-opt optimization for various numbers of layers $\in [2, 12]$ and evolution times $t \in \{0.5, 0.6, 0.75, 0.9\}$ on the one dimensional TFIM (at transverse field strength $g=3$) for eight qubits with periodic boundaries. We investigate the depth vs. $\epsilon$ scaling for each $t$ value and plot these data points in Fig. \ref{fig:rqc_opt_numerical}. When we scale the y axis to plot $t \log(t/\epsilon)$ for each of the data points, we observe that the plots of different evolution times overlap well.

We assume the scaling can be fit to a function of form:
\begin{equation}
\label{eq:fit}
    c_0 t^{c_1} \log(t/\epsilon) + c_2 \log(1/\epsilon) + c_3
\end{equation}
and we retrieve optimal parameters $\{c_i\}_i$ for the data points we plot in Fig. \ref{fig:rqc_opt_numerical}. We report these parameters to be: (0.192, 1.083, 0.367, -1.57). This scaling fit is consistent with the upper bound given in Theorem \ref{thm:existence2}, for fixed $N$, as $\log$ is a special instance of $\text{polylog}$ functions.  We compute 95\% confidence intervals using the Jacobian-based covariance estimate of the nonlinear regression, yielding the uncertainty of $\pm 0.087$ for $c_1$.

We also justify our claim of "near-optimality", through formalizing the lower scaling bound with respect to accuracy. Any procedure to distinguish two evolutions up to spectral norm $\epsilon$ will need precision costs scaling with $\log(1/\epsilon)$ \cite{Atia_2017}. Combined in an additive way, we arrive at the lower bound for the "query complexity":
\begin{equation}
\label{eq:query_comp_lower_bound}
    \Omega(t + \log(1/\epsilon))
\end{equation}
which was achieved by Quantum Signal Processing framework, as proposed by Low \& Chuang \cite{Low_2017}. Their proposal assumes, that one has access to an oracle representation of the Hamiltonian (e.g. through block encoding) and shows the number of queries made to this oracle saturates the lower bound in Eq. \ref{eq:query_comp_lower_bound}. This bound can also be applied to gate complexity if one can encode the oracle exactly with the given gate model. However, such an assumption of having access to an oracle that encodes the Hamiltonian up to arbitrary precision cannot be fulfilled in general, unless the Hamiltonian assumes favorable structure (e.g. linear combination of unitaries, QROM form) \cite{camps2023explicitquantumcircuitsblock, qsvt}. In general, implementing such an oracle up to precision $\epsilon$ brings additional computational cost that scales with $\text{polylog}(1/\epsilon)$ \cite{Clader_2022}. Hence, one can formalize a generic lower bound for the "gate complexity" (or circuit depth) of encoding the time evolution operator with simulation time $t$ and accuracy $\epsilon$ as:
\begin{equation}
\label{eq:gate_complexity_lower_bound}
    \Omega(t \hspace{0.1cm} \text{polylog}(1/\epsilon))
\end{equation}
in the $t \gg \log(1/\epsilon)$ limit.

\begin{figure}
    \centering
    \includegraphics[width=0.99\linewidth]{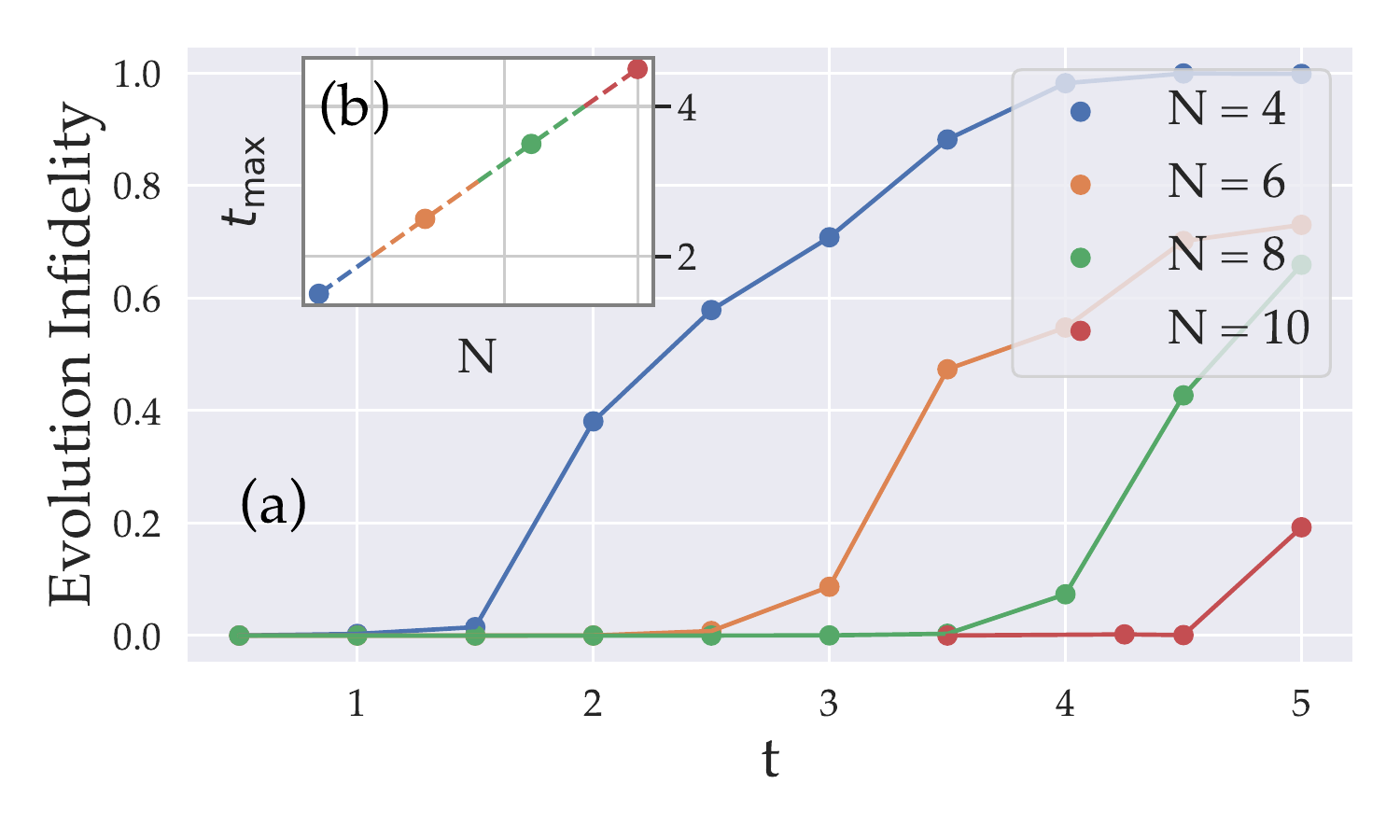}
    \caption{\textbf{(a)} Evolution infidelity (\ref{eq:ev_infidelity}) vs. evolution time of the RQC-opt results for the one dimensional TFIM, g=3 (\ref{eq:TFIM}), run with various Ansatz sizes $N$ and evaluated on $12$ qubits. \textbf{(b)} Inferred maximal evolution time $t_{\text{max}}$ with respect to Ansatz size N. We take the largest evolution time (among the ones considered) with infidelity below 0.05 (an empirical threshold) as the corresponding $t_{\text{max}}$. We observe the conjectured linear scaling (Eq. \ref{eq:tmax_scaling}) with respect to Ansatz size $N$.}
    \label{fig:tmax}
\end{figure}

Additionally, we verify the scaling of the maximal evolution time $t_{\text{max}}$ (for which the transferability of the optimized gates holds) through numerical simulations. In Sec. \ref{sec:sec2}, we argue this scaling to be linear (for $D=1$) in Ansatz size $N$, through the light cone shaped information propagation mechanism. To verify this, we run RQC-opt with a 15 layers Ansatz, different evolution times $t \in [0.1, 5]$ and different Ansatz system sizes $N \in \{4, 6, 8, 10\}$. We then evaluate these optimization results in terms of accuracy of approximating the time evolution of a 12 qubit system. We assume that 15 layered Ansätze are expressive enough for all evolution times considered and plot the evolution infidelity (\ref{eq:ev_infidelity}) vs. evolution time for different Ansatz sizes in Fig. \ref{fig:tmax}.a. We observe that the evolution infidelity stays negligibly small until a cut-off time $t_{\text{max}}$ and then starts increasing, for all the Ansatz sizes considered. We then plot these empirical cut-off values with respect to the Ansatz sizes in Fig. \ref{fig:tmax}.b and observe the expected scaling behavior given in Eq. \ref{eq:tmax_scaling}. These simulations were run for the one dimensional TFIM (at transverse field strength $g=3$) with periodic boundary conditions.

\section{Circuit Equivalence for QFT based QPE}
\label{sec:A4}
In the textbook standard, Quantum Fourier Transform (QFT) appended QPE variant with multiple ancillae and controlled evolutions, the collective state of ancillae after controlled operations is:
\begin{equation}
    \frac{1}{\sqrt{2^n}} \sum_{x=0}^{2^n-1} e^{-i \lambda t_0 x} \ket{x}
\end{equation}
where $n$ is the number ancillae, $\lambda$ is the eigenvalue of the manifold one is simulating and $t_0$ is the basis time step of the QPE protocol. This expression is equivalent to the following, up to the global phase $e^{i \lambda(2^n-1)t_0/2}$:
\begin{equation}
    \frac{1}{\sqrt{2^n}} \sum_{x=0}^{2^n-1} e^{i \lambda \frac{t_0}{2}  (2^n-1-x)} e^{-i \lambda \frac{t_0}{2} x} \ket{x}
\end{equation}
which effectively corresponds to replacing each controlled time evolution unitary (indexed $k$) with a control sequence where ancilla controls the evolution direction of a time evolution operator (similar to the expression given in Eq. \ref{eq:cU_equivalence}) with evolution time set to $t_02^{k-1}$ for $k=0, \dots, n-1$.

\section{Two dimensional Ansatz circuits for TICC}
\label{sec:A2}
As mentioned in the main text, we generalize the Ansatz construction proposed in \cite{rqcopt1} to two dimensional systems. In this case, the Ansatz circuits do not assume a clean brickwall layout where connectivity permutations alternate between sites $[(0, 1), (2, 3), \dots]$ and $[(N-1, 0), (1, 2), \dots]$ but have an increased number of permutations. We extract these permutations for the two dimensional square and triangular lattice geometries, using the adjacency matrices of the corresponding geometry and system size. We take these adjacency matrices from the python library \texttt{qib}, hence we use the same definition for the triangular and square lattice geometries, employed there. 

\begin{figure}
    \centering
    \includegraphics[width=0.99\linewidth]{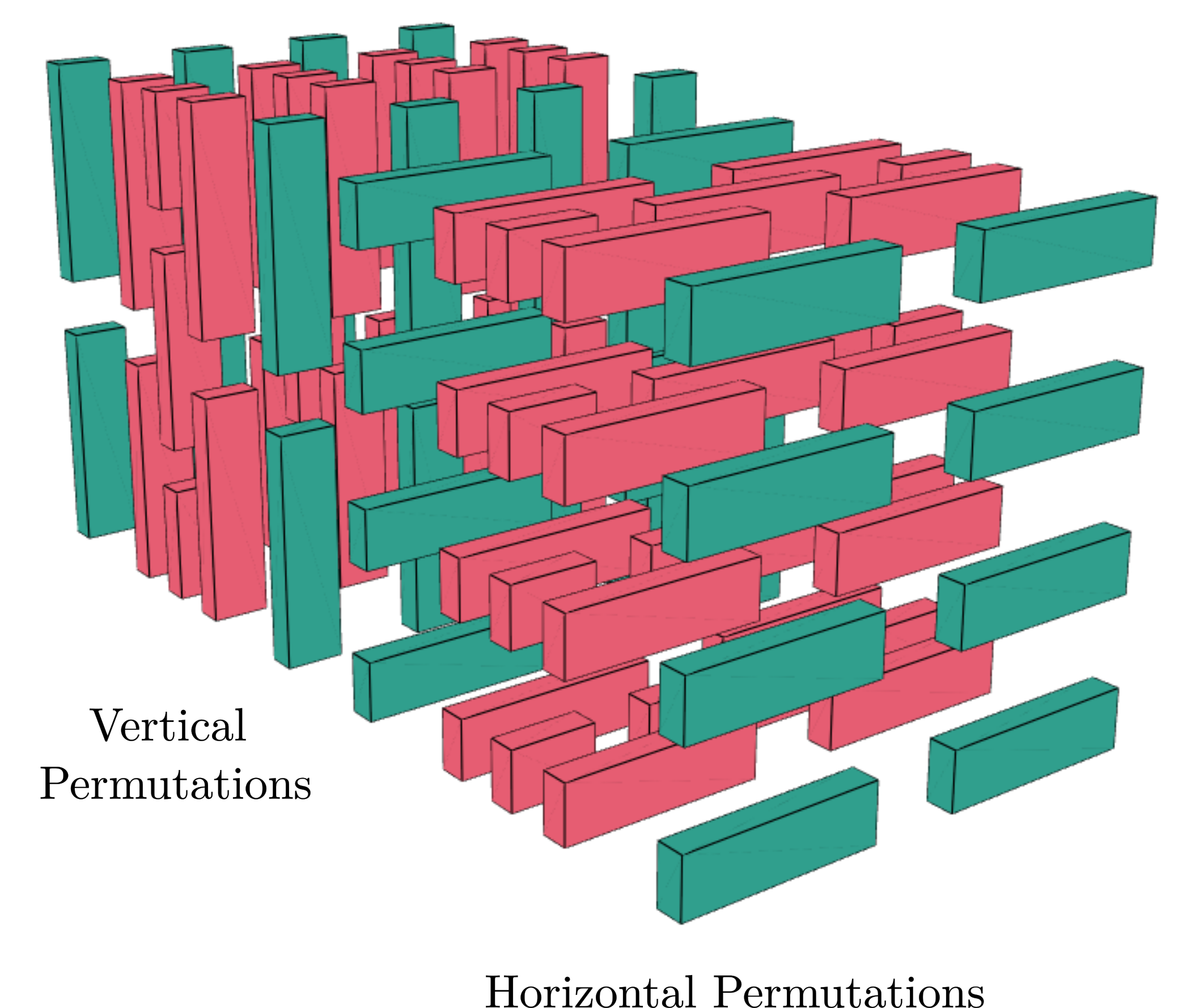}
    \caption{Two dimensional TICC circuit Ansatz of 10 layers, employed to capture the connectivity of a 4$\times$4 square lattice with nearest neighbor interactions. Green gates  correspond to the control layers, which flip the evolution direction. 5 layered block in the back encode the vertical permutations of the square lattice, whereas the remaining 5 layers in the in the front encode the horizontal permutations.}
    \label{fig:fig7}
\end{figure}

In Fig. \ref{fig:fig7}, we visualize an example TICC Ansatz circuit which can be used to approximate the controlled time evolution of a system on the 4$\times$4 square lattice geometry. We use this particular circuit construction for the reported TICC results of the square lattice TFIM.

In the following, we explicitly give the inequivalent classes of nearest neighbour permutations for the considered geometries. We use:
\begin{multline}
    \text{perms}_1 = \\ [(0, 4, 1, 5, 2, 6, 3, 7, 8, 12, 9, 13, 10, 14, 11, 15), \\ (4, 8, 5, 9, 6, 10, 7, 11, 12, 0, 13, 1, 14, 2, 15, 3)],
\end{multline}
\begin{multline}
    \text{perms}_2 = \\ [(0, 1, 4, 5, 8, 9, 12, 13, 2, 3, 6, 7, 10, 11, 14, 15), \\
  (1, 2, 5, 6, 9, 10, 13, 14, 3, 0, 7, 4, 11, 8, 15, 12)]
\end{multline}
for the 4$\times$4 square lattice (number of inequivalent classes $d=2$ as mentioned in Sec. \ref{sec:sec4}) and
\begin{multline}
    \text{perms}_1 = \\ [(0, 1, 2, 3, 4, 5, 6, 7, 8, 9, 10, 11, 12, 13, 14, 15), \\(1, 2, 3, 0, 5, 6, 7, 4, 9, 10, 11, 8, 13, 14, 15, 12)]
\end{multline}
\begin{multline}
    \text{perms}_2 = \\ [(0, 5, 10, 15, 3, 4, 9, 14, 2, 7, 8, 13, 1, 6, 11, 12), \\ (5, 10, 15, 0, 4, 9, 14, 3, 7, 8, 13, 2, 6, 11, 12, 1)] 
\end{multline}
\begin{multline}
    \text{perms}_3 = \\ [(0, 4, 8, 12, 1, 5, 9, 13, 2, 6, 10, 14, 3, 7, 11, 15), \\(4, 8, 12, 0, 5, 9, 13, 1, 6, 10, 14, 2, 7, 11, 15, 3)]
\end{multline}
for the 4$\times$4 triangular lattice (number of inequivalent classes $d=3$). In this notation, each pair of consecutive entries $(a_{2j}, a_{2j+1})$ within a permutation denotes a nearest-neighbor bond. Each such permutation corresponds to one circuit layer, because two qubit gate operations that connect each neighbor of the given permutation can be parallelized. 

\section{Phase Estimation Simulations with TICC}
\label{sec:A_QPE}
Here, we present the iterative QPE protocol which we used to obtain the energy estimation results plotted in Fig. \ref{fig:fig4}.a. This iterative protocol has been studied and analyzed in depth in \cite{rpe}. In this algorithm, one starts with an initial guess $E_{0}$ for the energy value and refine this guess digit by digit by doubling the evolution time $t=2^k$ at each iteration $k$. Depending on the initial absolute distance of the guessed value from the true energy, one starts the iterative protocol from smaller $t$ values. The maximal precision one can reach in this iterative process is ultimately limited by the hardware resources and coherence, as increasing $t$ translates to increasing circuit depths for each circuit.

Emulator results we provide in Fig. \ref{fig:fig4}.a were run for $k$ values -3 and -2. For each stage of this iterative protocol, we run phase estimations for multiple, tightly packed $t$ values ($t \in \{0.122, 0.124, 0.126, 0.128\}$ for $k=-3$ and $t \in \{0.22, 0.24, 0.26, 0.28\}$ for $k=-2$). Then for each stage, we fit the estimated phases into a phase curve, whose fit function is $\cos\left(E^{(k)}_{\text{fit}} t\right)$ for the real part and $\sin\left(E^{(k)}_{\text{fit}} t\right)$ for the imaginary part of the phase. This idea, inspired by the authors of \cite{Ding_2023}, refines the phase estimate significantly while causing tolerable increase in required hardware resources. Crucially it does not increase circuit depths but increases total hardware run time. We visualize this fitting idea in the bottom plot of Fig. \ref{fig:fig1}.c.

\begin{figure}
    \centering
    \includegraphics[width=0.9\linewidth]{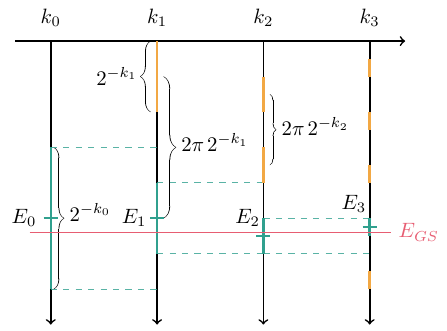}
    \caption{Illustration of iterative energy estimation protocol for three stages. At each iteration $k=k_j, \hspace{0.1cm} j\in\{1, 2, 3\}$; phase $e^{-i 2^k E_{GS}}$ is estimated with a Hadamard test. Then the $2\pi 2^{-k}$ periodic translations of the phase angle is considered (orange), whose guaranteed precision is $2^{-k-1}$ at each stage $k$. Among these translations, the one closest to the previous estimate $E_{j-1}$ is chosen as the new estimate $E_j$ (cyan).}
    \label{fig:fig5}
\end{figure}

After acquiring this refined phase estimate, we use the protocol described in \cite{rpe} to obtain the energy estimate of the current stage $E_k$ from the previous estimate $E_{k-1}$. This technique, visualized in Fig. \ref{fig:fig5} and summarized in Alg. \ref{alg:alg3}, assumes that the energy estimate $E_{k-1}$ from stage $k-1$ satisfies $|E_{\text{GS}} - 
E_{k-1}| \leq 2^{-k}$ (where $E_{\text{GS}}$ is the true ground state energy) and chooses $E_k$ from the periodic, $2\pi 2^{-k}$ translations of the phase angle $E^{(k)}_{\text{fit}}$, that is closest to the previous estimate $E_{k-1}$.

We also formalize an error bar calculation protocol in this work. We integrate a randomized Pauli twirling scheme to our compilation. This compilation trick allows us to treat a wide range of hardware noise channels as a global depolarizing channel \cite{Silva_2008,  Geller_2013, Wallman_2016, Liu_PRA}, whose effect on expectation value measurements is an exponential damping towards zero. This means that increasing hardware noise causes both the real and the imaginary part of the estimated phase to near 0, effectively causing amplitude dampening. For the aforementioned iterative energy estimation protocol, this amplitude dampening effect is inherently ignored, as what determines the angle of the estimated phase is only the ratio between the real and imaginary parts. 

To incorporate this amplitude reducing effect of the hardware noise into our error bars, we propose to amplify the statistical, standard deviation of the expectation value measurement by the inverse of the measured phase amplitude - effectively projecting the estimated phase onto the unit circle in the complex plane. This process is visualized in the top plot of Fig. \ref{fig:fig1}.c. These error bars are then incorporated into the fitting step (visualized in the bottom plot of Fig. \ref{fig:fig1}.c), where data uncertainties of the phase estimates are propagated into parameter uncertainties (in this case, the parameter of the fit corresponds to our refined energy estimate $E^{(k)}_{\text{fit}}$), using \texttt{curve\_fit} function of the python library \texttt{scipy}. Uncertainty estimate for the fitted parameter $E^{(k)}_{\text{fit}}$ is obtained from the inverse of the Fisher information matrix, returned by the \texttt{curve\_fit} function.

\begin{algorithm}
\caption{Iterative Phase Estimation (as proposed by Ni et al.~\cite{rpe})}
\label{alg:alg3}
\KwData{$\ket{\psi}, \; E_0,\; (k_0, k_1, \dots k_J)$}
\For {$k = k_1, \dots , k_J$}{
    $E^{(k)}_{\text{fit}} \approx \mathrm{arg}\, \braket{\psi|e^{i2^kH}|\psi}$ 
    \tcp{Hadamard test and curve fit (Fig.~\ref{fig:fig1}.b \& \ref{fig:fig1}.c)}
    $S_k \coloneqq \left\{ \tfrac{2j\pi + E^{(k)}_{\text{fit}}}{2^k} \;\middle|\; j=0,\dots,2^k-1 \right\}$
    
    $E_k = \operatorname*{argmin}_{E \in S_k} \Big[ \pi - \big| (E - E_{k-1} \bmod 2\pi) - \pi \big| \Big]$
}
\Return $E_J$ \tcp{Energy estimate modulo $2\pi$}
\end{algorithm}

For the iterative energy estimation process, we also determine a termination condition , related to the highest $k^{\text{th}}$ stage that can be successfully terminated, given the hardware noise limitations. A successful approximation $E_k$ at stage $k$, should satisfy: $|E_k - E_{\text{GS}}| \leq 2^{-k-1}$ (in other words, $k^{\text{th}}$ binary digits should match). If the extracted error bars for the given stage $k$ exceeds $2^{-k-1}$, we determine that stage to be unsuccessful and terminate the iterative process. For our aforementioned emulations under realistic hardware noise levels, this maximal $k$ stage corresponds to $k=-2$.

In the emulations reported in this work, we focused on the high field regime of TFIM, where the true ground state is sufficiently well approximated for the purposes of iterative QPE, by the ground state of TFIM at $g \rightarrow \infty$ limit, hence we use this state as the initial state of system qubits. As shown in \cite{rpe}, this iterative QPE protocol requires the initial state of system qubits to have at least around 53\% overlap with the ground state. This aforementioned initial state is easy to prepare with just single qubit operations on the system qubits and delivers high enough overlap with the true ground state: \{58\%, 69\%, 76.5\%, 81.6\%\}, for the field strengths considered $g \in \{1.5, 2, 2.5, 3\}$. Hence, we initialize the system qubits in this state for our QPE protocols.

\begin{table}
    \centering
    \includegraphics[width=1\linewidth]{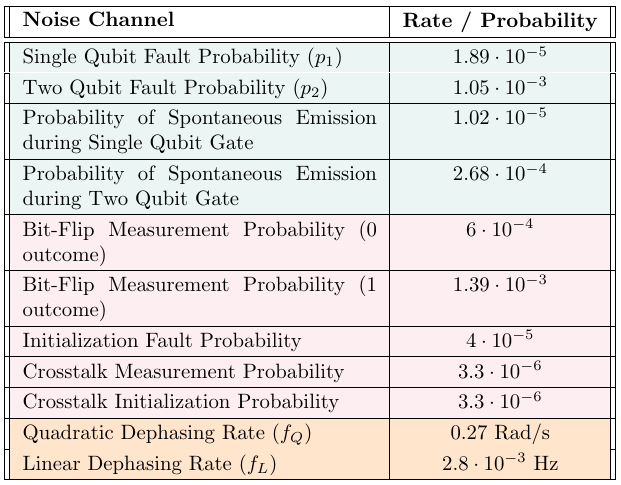}
    \caption{Noise parameters of the Quantinuum H2 trapped ion quantum computer, as reported in \cite{Anserson_2021, QuantinuumH2Emulators2025}.}
    \label{tab:tab2}
\end{table}

\section{Noise model of Emulator Simulations}
\label{sec:A3}
For the QPE results we report in Sec. \ref{sec:sec4}, we use the noise aware emulator of the Quantinuum H2 trapped ion quantum computer, accessed through the "Quantinuum Nexus" platform.

This emulator uses state vector simulations and models the hardware noise of the trapped ion quantum computer as a combination of local, asymmetric depolarizing channels for one and two qubit operations, with a non-zero probability of spontaneous emission during the operation. They model a spontaneous emission event, as applying Pauli-X with 1/4, Pauli-Y with 1/4 and applying a leakage event with 1/2 probability. Fault probability for one qubit operation is fixed and given in Table \ref{tab:tab2} as $p_1$. Fault probability of the hardware native ZZPhase gate depends on the geometric phase angle parameter $\Theta$ of the gate and is given as:
\begin{equation}
    \left( 1.52 \frac{|\Theta|}{\pi} + 0.24 \right) p_2
\end{equation}
with $p_2$ reported in Table \ref{tab:tab2}. The probabilities of spontaneous emission during a one and two qubit gate, are also listed in Table \ref{tab:tab2}.

In this noise model, initialization fault and measurement bit flip (for the measurement of 0 and 1) probabilities are also explicitly incorporated. Moreover, crosstalk probability between qubits resulting in initialization and measurement fault are separately considered. Probability rates of all these noise channels are listed in Table \ref{tab:tab2}.

Additional to fault probabilities during gate operations, initialization and readout; noise model also incorporates the dephasing noise that qubits are subjected to due to idling and transport. This noise channel is a combination of a coherent, quadratic and a linear dephasing channel. The former is applied as an RZ rotation with angle $f_Q \cdot \tau$ and the latter is applied as a single phase flip (applying Pauli-Z) operation with probability $f_L \cdot \tau$, for the idling/transport time $\tau$. These parameters $f_Q, f_L$ are also both listed in Table \ref{tab:tab2}.

All of the noise parameters listed in this section, are publicly available from Quantinuum’s website; no proprietary data were used.

\nocite{erenaykrcn_ccU_2026}
\bibliography{bibliography.bib}
\end{document}